\documentclass[onecolumn,draftcls,journal,10p]{IEEEtran}
\usepackage[T1]{fontenc}
\usepackage[latin9]{inputenc}
\usepackage{geometry}
\usepackage{color}
\usepackage{float}
\usepackage{amsmath}
\usepackage{amsthm}
\usepackage{amssymb}
\usepackage{graphicx}
\usepackage[unicode=true,
 bookmarks=true,bookmarksnumbered=true,bookmarksopen=true,bookmarksopenlevel=1,
 breaklinks=false,pdfborder={0 0 0},pdfborderstyle={},backref=false,colorlinks=false]
 {hyperref}
\hypersetup{pdftitle={Your Title},
 pdfauthor={Your Name},
 pdfpagelayout=OneColumn, pdfnewwindow=true, pdfstartview=XYZ, plainpages=false}

\makeatletter
\theoremstyle{plain}
\newtheorem{thm}{\protect\theoremname}
\theoremstyle{plain}
\newtheorem{lem}[thm]{\protect\lemmaname}
\theoremstyle{definition}
\newtheorem{defn}[thm]{\protect\definitionname}
\theoremstyle{remark}
\newtheorem{rem}[thm]{\protect\remarkname}

\usepackage{epstopdf}

\usepackage{tikz}
\usetikzlibrary{arrows}
\usetikzlibrary{shadows}

\usetikzlibrary{shapes}

\usetikzlibrary{arrows,decorations.pathmorphing}

\usepackage{hyperref}
\@ifundefined{rangeHsb}{\usepackage{xcolor}}{}
\usepackage{pdfsync}

\makeatother

\providecommand{\definitionname}{Definition}
\providecommand{\lemmaname}{Lemma}
\providecommand{\remarkname}{Remark}
\providecommand{\theoremname}{Theorem}

\begin{document}
\title{Consensus on Matrix-weighted \\Time-varying Networks}
\author{Lulu Pan, Haibin~Shao,~\IEEEmembership{Member,~IEEE,} Mehran~Mesbahi,~\IEEEmembership{Fellow,~IEEE,}\\
Yugeng~Xi,~\IEEEmembership{Senior Member,~IEEE}, Dewei Li\thanks{This work is supported by the National Science Foundation of China
(Grant No. 61973214, 61590924, 61963030) and Natural Science Foundation
of Shanghai (Grant No. 19ZR1476200) and in part by the U.S. 
Air Force Office of Scientific Research (Grant No. FA9550-16-1-0022). (Corresponding
author: Haibin Shao)}\thanks{Lulu Pan, Haibin Shao, Yugeng Xi and Dewei Li are with the Department
of Automation and the Key Laboratory of System Control and Information
Processing Ministry of Education of China, Shanghai Jiao Tong University,
Shanghai 200240, China (\{llpan,shore,ygxi,dwli\}@sjtu.edu.cn). Mehran
Mesbahi is with the Department of Aeronautics and Astronautics, University
of Washington, Seattle, WA, 98195-2400, USA (mesbahi@uw.edu).}}
\maketitle
\begin{abstract}
This paper examines the consensus problem on time-varying matrix-weighed
undirected networks. First, we introduce the matrix-weighted integral network 
for the analysis of such networks.
Under mild assumptions on the switching pattern of the time-varying
network, necessary and/or sufficient conditions for which average consensus
can be achieved are then provided in terms of the null space of matrix-valued
Laplacian of the corresponding integral network. In particular, for periodic
matrix-weighted time-varying networks, necessary
and sufficient conditions for reaching average consensus is obtained from an
algebraic perspective. Moreover, we show that if the integral network with period $T >0$
has a positive spanning tree over the time span $[0,T)$,
average consensus for the node states is achieved. 
Simulation results are provided to demonstrate the theoretical analysis.
\end{abstract}

\section{Introduction}

Reaching consensus is an important construct in distributed
coordination of multi-agent systems \cite{mesbahi2010graph,olfati2007consensus,zhang2018fully,degroot1974reaching}.
Although the consensus problem has been extensively investigated in the literature,
it has often been assumed that the network has scalar-weighted edges;
extensions of the scalar weights to matrix-valued weights
has become relevant in order to characterize interdependencies
among multi-dimensional states of neighboring agents. 
Recently, a broader category of networks referred to as matrix-weighted
networks has been introduced to address such interdependencies~\cite{trinh2018matrix,sun2018dimensional}.
In fact, matrix-weighted networks arise in scenarios such as graph
effective resistance examined in the context of 
distributed control and estimation \cite{barooah2006graph,tuna2017observability},
logical inter-dependencies amongst topics in opinion evolution \cite{friedkin2016network,ye2018continuous},
bearing-based formation control \cite{zhao2015translational}, dynamics of an array
of coupled LC oscillators \cite{tuna2019synchronization}, as well
as consensus and synchronization on matrix-weighted networks \cite{trinh2018matrix,tuna2016synchronization,pan2018bipartite}.

For matrix-weighted networks, network connectivity does not translates to
achieving consensus. To this end, properties of weight matrices play an important role in characterizing
consensus. For instance, positive
definiteness and positive semi-definiteness of weight matrices have been 
employed to provide consensus conditions in \cite{trinh2018matrix};
negative definiteness and negative semi-definiteness of weight matrices
are further introduced in \cite{pan2018bipartite,su2019bipartite}.
In the meantime, the notion of network connectivity
can be further extended for matrix-valued networks. For instance, one
can identify edges with positive/negative definite matrices as ``strong'' connections; whereas
an edge weighted by positive/negative semi-definite matrices can be considered
as a ``weak'' connection \cite{trinh2017theory}.

To the best of our knowledge, conditions under which consensus
can be achieved for time-varying matrix-weighted networks have
not been developed in the literature; this is in contrast with conditions 
that have been examined for scalar-weighted networks \cite{cao2008reaching,olfati2004consensus,ren2005consensus,Jadbabaie2003,cao2011necessary,moreau2005stability,meng2018uniform,meng2014modulus,proskurnikov2014consensus}.
In this paper, we provide necessary and/or sufficient conditions for
achieving consensus on matrix-weighed time-varying  networks. Under
mild assumptions on the switching pattern for such networks,
necessary and/or sufficient conditions for which average consensus
is achieved are provided in terms of the null space of the matrix-valued
Laplacian of the associated integral networks. In particular, for periodic
matrix-weighted time-varying networks with period $T>0$, a necessary
and sufficient condition for average consensus is obtained; we
further show that from a graph-theoretic perspective, when the integral
network over time span $[0,T)$ has a positive spanning tree, then 
average consensus is achieved. Simulation results are provided to demonstrate
the theoretical analysis.

The remainder of this paper is organized as follows. Preliminaries
are introduced in \S\ref{sec:Preliminaries}. The problem formulation
is provided in \S\ref{sec:Problem-Formulation}, followed
by the consensus conditions in \S\ref{sec:Consensus-on-General}
and \S\ref{sec:Consensus-on-Periodic}, respectively. A simulation example is
presented in \S\ref{sec:Simulation-Results} followed by concluding
remarks in \S~\ref{sec:Conclusion}.

\section{Preliminaries \label{sec:Preliminaries}}

Let $\mathbb{R}$, $\mathbb{N}$ and $\mathbb{Z}_{+}$ be the set
of real numbers, natural numbers and positive integers, respectively.
For $n\in\mathbb{Z}_{+}$, denote $\underline{n}=\left\{ 1,2,\ldots,n\right\}$.
A symmetric matrix $M\in\mathbb{R}^{n\times n}$ is positive definite,
denoted by $M\succ0$, if $\boldsymbol{z}^{\top}M\boldsymbol{z}>0$
for all $\boldsymbol{z}\in\mathbb{\mathbb{R}}^{n}$ and $\boldsymbol{z\not}=\boldsymbol{0}$
and is positive semi-definite, denoted by $M\succeq0$, if $\boldsymbol{z}^{\top}M\boldsymbol{z}\ge0$
for all $\boldsymbol{z}\in\mathbb{\mathbb{R}}^{n}$.
The null space of a matrix $M\in\mathbb{R}^{n\times n}$ is denoted by $\text{{\bf null}}(M)=\left\{ \boldsymbol{z}\in\mathbb{R}^{n}|M\boldsymbol{z}=\boldsymbol{0}\right\} $. 
\begin{lem}
\label{lem:Rayleigh Theorem}\cite{horn2012matrix} Let $M\in\mathbb{R}^{n\times n}$
be symmetric with eigenvalues $\lambda_{1}\leq\cdots\leq\lambda_{n}$.
Let $\boldsymbol{x}_{i_{1}},\cdots,\boldsymbol{x}_{i_{k}}$ be mutually
orthonormal vectors such that $M\boldsymbol{x}_{i_{p}}=\lambda_{i_{p}}\boldsymbol{x}_{i_{p}}$,
where $i_{p}\in\mathbb{Z}_{+}$, $p\in\underline{k}$ and $1\leq i_{1}<\cdots<i_{k}\leq n$.
Then
\[
\lambda_{i_{1}}=\underset{\{\boldsymbol{x}\not={\bf 0},\boldsymbol{x}\in S_k\}}{\text{{\bf min}}}\frac{\boldsymbol{x}^{\top}M\boldsymbol{x}}{\boldsymbol{x}^{\top}\boldsymbol{x}},
\]
and
\[
\lambda_{i_{k}}=\underset{\{\boldsymbol{x}\not={\bf 0},\boldsymbol{x}\in S_k\}}{\text{{\bf max}}}\frac{\boldsymbol{x}^{\top}M\boldsymbol{x}}{\boldsymbol{x}^{\top}\boldsymbol{x}},
\]
where $S_k=\text{{\bf span}}\{\boldsymbol{x}_{i_{1}},\cdots,\boldsymbol{x}_{i_{k}}\}$.
\end{lem}

\section{Problem Formulation \label{sec:Problem-Formulation}}

Consider a multi-agent system consisting of $n>1$ ($n\in\mathbb{Z}_{+}$)
agents whose interaction network is characterized by a matrix-weighted
time-varying graph $\mathcal{G}(t)=(\mathcal{V},\mathcal{E}(t),A(t))$, where $t$ refers to the time index. The node and edge sets of
$\mathcal{G}$ are denoted by $\mathcal{V}=\left\{ 1,2,\ldots,n\right\} $
and $\mathcal{E}(t)\subseteq\mathcal{V}\times\mathcal{V}$, respectively.
The weight on the edge $(i,j)\in\mathcal{E}(t)$ is encoded by the symmetric
matrix $A_{ij}(t)\in\mathbb{R}^{d\times d}$ such that $A_{ij}(t)\succeq0$
or $A_{ij}(t)\succ0$, and $A_{ij}(t)=0_{d\times d}$ for $(i,j)\not\in\mathcal{E}(t)$.\textbf{
}Thereby, the matrix-weighted adjacency matrix $A(t)=[A_{ij}(t)]\in\mathbb{R}^{dn\times dn}$
is a block matrix such that the block located in its $i$-th row and
the $j$-th column is $A_{ij}(t)$. It is assumed that $A_{ij}(t)=A_{ji}(t)$
for all $i\not\not=j\in\mathcal{V}$ and $A_{ii}(t)=0_{d\times d}$
for all $i\in\mathcal{V}$. 

Denote the state of an agent $i\in\mathcal{V}$ as $\boldsymbol{x}_{i}(t)=[x_{i1}(t),x_{i2}(t),\ldots,x_{id}(t)]^{\top}\in\mathbb{R}^{d}$ evolving according to the protocol,
\begin{equation}
\dot{\boldsymbol{x}}_{i}(t)=-\sum_{j\in\mathcal{N}_{i}(t)}A_{ij}(t)(\boldsymbol{x}_{i}(t)-\boldsymbol{x}_{j}(t)),\thinspace i\in\mathcal{V},\label{equ:matrix-consensus-protocol}
\end{equation}
where $\mathcal{N}_{i}(t)=\left\{ j\in\mathcal{V}\,|\,(i,j)\in\mathcal{E}(t)\right\} $
denotes the neighbor set of agent $i\in\mathcal{V}$ at time $t$.
Note that protocol \eqref{equ:matrix-consensus-protocol} degenerates
into the scalar-weighted case when $A_{ij}(t)=a_{ij}(t)I$, where $a_{ij}(t)\in\mathbb{R}$
and $I$ denotes the $d\times d$ identity matrix. 

Let $C(t)=\text{{\bf diag}}\left\{ C_{1}(t),C_{1}(t),\cdots,C_{n}(t)\right\} \in\mathbb{R}^{dn}$
be the matrix-valued degree matrix of $\mathcal{G}(t)$,
where $C_{i}(t)=\sum_{j\in\mathcal{N}_{i}}A_{ij}(t)\in\mathbb{R}^{d\times d}$.
The matrix-valued Laplacian is subsequently defined as $L(t)=C(t)-A(t)$.
The dynamics of the overall multi-agent system now admits the form,
\begin{equation}
\dot{\boldsymbol{x}}(t)=-L(t)\boldsymbol{x}(t),\label{equ:matrix-consensus-overall}
\end{equation}
where $\boldsymbol{x}(t)=[\boldsymbol{x}_{1}^{\top}(t),\boldsymbol{x}_{2}^{\top}(t),\ldots,\boldsymbol{x}_{n}^{\top}(t)]^{\top}\in\mathbb{R}^{dn}$.
\begin{defn}
Let $\boldsymbol{x}_{f}=\boldsymbol{1_{n}}\otimes(\frac{1}{n}\sum_{i=1}^{n}\boldsymbol{x_{i}}(0))$.
Then the multi-agent system \eqref{equ:matrix-consensus-overall}
admits an average consensus solution if $\lim{}_{t\rightarrow\infty}\boldsymbol{x}_{i}(t)=\lim{}_{t\rightarrow\infty}\boldsymbol{x}_{j}(t)=\boldsymbol{x}_{f}$
for all $i,j\in\mathcal{V}$.
\end{defn}

This work aims to investigate the necessary and/or sufficient
conditions under which the multi-agent system \eqref{equ:matrix-consensus-overall}
admits an average consensus solution. It is well-known that network
connectivity plays a central role in determining consensus for scalar-weighted
networks \cite{olfati2004consensus}. 
However, as we shall show subsequently, definiteness
of the weight matrices is also a crucial factor in examining consensus for a matrix-weighted
networks in addition to its connectivity.
First, we shall recall a few facts on network connectivity. In graph theory, network
connectivity captures how a pair of nodes in the network can be ``connected'' by traversing
a sequence of consecutive edges called paths. A path of $\mathcal{G}(t)$ is a sequence
of edges of the form $(i_{1},i_{2}),(i_{2},i_{3}),\ldots,(i_{p-1},i_{p})$,
where nodes $i_{1},i_{2},\ldots,i_{p}\in\mathcal{V}$ are distinct;
in this case we say that node $i_{p}$ is reachable from $i_{1}$. 
The graph $\mathcal{G}(t)$ is connected if any two distinct nodes in $\mathcal{G}(t)$
are reachable from each other. A tree is a connected graph with $n\ge2$
nodes and $n-1$ edges where $n\in\mathbb{Z}_{+}$. 
For matrix weighted graphs, we adopt the following terminology.
An edge $(i,j)\in\mathcal{E}(t)$ is positive definite or positive semi-definite
if the associated weight matrix $A_{ij}(t)$ is positive definite
or positive semi-definite, respectively. A positive path in $\mathcal{G}(t)$
is a path such that every edge on this path is positive definite.
A tree in $\mathcal{G}(t)$ is a positive tree if every edge contained
in this tree is positive definite. A positive spanning tree of $\mathcal{G}(t)$
is a positive tree containing all nodes in $\mathcal{G}(t)$. 
\section{Consensus on General Matrix-weighted Time-varying Networks \label{sec:Consensus-on-General}}

In order to analyze multi-agent systems of the form~\eqref{equ:matrix-consensus-overall},
we adopt the following assumption on the matrix-weighted time-varying
network~\cite{olfati2004consensus,ren2005consensus,cao2011necessary,meng2018uniform}.

\textbf{Assumption 1. }There exists a sequence $\{t_{k}|k\in\mathbb{N}\}$
such that $\lim_{k\rightarrow\infty}t_{k}=\infty$ and $\triangle t_{k}=t_{k+1}-t_{k}\in[\alpha,\beta]$
for all $k\in\mathbb{N}$, where $\beta>\alpha>0$, $t_{0}=0$, and
$\mathcal{G}(t)$ is time-invariant for $t\in[t_{k},t_{k+1})$ for
all $k\in\mathbb{N}$.


When $L(t)=L$ for all $t\in[0,\infty)$, then \eqref{equ:matrix-consensus-overall}
encodes the consensus protocol on a time-invariant network. The
following observation characterizes the structure of the null space of matrix-valued
Laplacian $L$ on time-invariant networks, that in turn, can determine the steady-state
of the network~\eqref{equ:matrix-consensus-overall}. 
\begin{lem}
\label{lem:1}\cite{trinh2018matrix} Let $\mathcal{G}=(\mathcal{V},\mathcal{E},A)$
be a matrix-weighted time-invariant network with matrix-valued Laplacian
$L$. Then $L\succeq0$ and $\text{{\bf null}}(L)=\text{{\bf span}}\left\{ \mathcal{R},\mathcal{H}\right\}$,
where $\mathcal{R}=\text{{\bf range}}\{\boldsymbol{1}\otimes I_{d}\}$
and
\begin{align*}
\mathcal{H=}\{&[\boldsymbol{v}_{1}^{T},\boldsymbol{v}_{2}^{T},\cdots,\boldsymbol{v}_{n}^{T}]^{T}\in\mathbb{R}^{dn}\mid\\
 & (\boldsymbol{v}_{i}-\boldsymbol{v}_{j})\in\text{{\bf null}}(A_{ij}),\,(i,j)\in\mathcal{E}\}.
\end{align*}
\end{lem}
Note that the null space of a matrix-valued Laplacian is not only determined
by the network connectivity, but also by the properties of weight
matrices; this is distinct from the scalar-weighted networks.
For matrix-weighted time-invariant networks, a condition under
which the multi-agent system \eqref{equ:matrix-consensus-overall}
achieves consensus is provided in the following lemma.
\begin{lem}
\label{lem:2}\cite{trinh2018matrix} Let $\mathcal{G}=(\mathcal{V},\mathcal{E},A)$
be a matrix-weighted time-invariant network with matrix-valued Laplacian
$L$. Then the multi-agent system \eqref{equ:matrix-consensus-overall}
admits an average consensus if and only if $\text{{\bf null}}(L)=\mathcal{R}$.
\end{lem}
\begin{defn}
Define the consensus subspace of the multi-agent system \eqref{equ:matrix-consensus-overall}
as $\mathcal{R}=\text{{\bf range}}\{\boldsymbol{1}\otimes I_{d}\}$.
\end{defn}
\begin{lem}
\label{lem:3}\cite{trinh2018matrix} Let $\mathcal{G}=(\mathcal{V},\mathcal{E},A)$
be a matrix-weighted time-invariant network. If $\mathcal{G}$ has
a positive spanning tree $\mathcal{T}$, then the network
\eqref{equ:matrix-consensus-overall} admits an average consensus.
\end{lem}
In order to characterize the related properties of the time-varying
networks $\mathcal{G}(t)$ over a given time span, we introduce the
notion of matrix-weighted integral network; this notion proves
crucial in characterizing algebraic and graph-theoretic conditions for reaching
consensus on matrix-weighted time-varying networks. 
\begin{defn}
\label{def:integral graph}Let $\mathcal{G}(t)=(\mathcal{V},\mathcal{E}(t),A(t))$
be a matrix-weighted time-varying network. Then the matrix-weighted
integral network of $\mathcal{G}(t)$ over time span $[\tau_{1},\tau_{2})\subseteq[0,\infty)$
is defined as $\widetilde{\mathcal{G}}_{[\tau_{1},\tau_{2})}=(\mathcal{V},\widetilde{\mathcal{E}},\widetilde{A})$,
where 
\[
\widetilde{A}=\frac{1}{\tau_{2}-\tau_{1}}\intop_{\tau_{1}}^{\tau_{2}}A(t)dt
\]
and
\[
\widetilde{\mathcal{E}}=\left\{ (i,j)\in\mathcal{V}\times\mathcal{V}\mid\intop_{\tau_{1}}^{\tau_{2}}A_{ij}(t)dt\succ0\; \text{or}\; \intop_{\tau_{1}}^{\tau_{2}}A_{ij}(t)dt\succeq0\right\} .
\]
\end{defn}
According to Definition \ref{def:integral graph}, denote by $\widetilde{C}$
as the matrix-weighted degree matrix of $\widetilde{\mathcal{G}}_{[\tau_{1},\tau_{2})}$, that is,
$\widetilde{C}=\frac{1}{\tau_{2}-\tau_{1}}\intop_{\tau_{1}}^{\tau_{2}}C(t)dt$.
Denote the matrix-valued Laplacian of $\widetilde{\mathcal{G}}_{[\tau_{1},\tau_{2})}$
as $\widetilde{L}_{[\tau_{1},\tau_{2})}$. Thus,

\begin{align*}
\widetilde{L}_{[\tau_{1},\tau_{2})} & =\widetilde{C}-\widetilde{A}=\frac{1}{\tau_{2}-\tau_{1}}\intop_{\tau_{1}}^{\tau_{2}}L(t).
\end{align*}

According to Assumption 1, we denote $\mathcal{G}(t)$ on dwell
time $t\in[t_{k},t_{k+1})$ as $\mathcal{G}_{[t_{k},t_{k+1})}(t)=\mathcal{G}^{k}$
and denote the associated matrix-valued Laplacian as $L^{k}$, where
$k\in\mathbb{N}$. The following lemma reveals the connection between
the null space of the matrix-valued Laplacian of a sequence of matrix-weighted
networks and that of the corresponding integral network.
\begin{lem}
\label{thm:null space equality} Let $\mathcal{G}(t)$ be a matrix-weighted
time-varying network satisfying \textbf{Assumption 1. }Then $\text{{\bf null}}(\widetilde{L}_{[t_{k^{\prime}},t_{k^{\prime\prime}})})=\mathcal{R}$
if and only if 
\[
\underset{i\in\underline{k^{\prime\prime}-k^{\prime}}}{\bigcap}\text{{\bf null}}(L^{k^{\prime}+i-1})=\mathcal{R},
\]
where $k^{\prime}<k^{\prime\prime}\in\mathbb{N}$.
\end{lem}
\begin{IEEEproof}
(Necessity) From the definition of matrix-valued Laplacian, one has
$\mathcal{R}\subseteq\underset{i\in\underline{k^{\prime\prime}-k^{\prime}}}{\bigcap}\text{{\bf null}}(L^{k^{\prime}+i-1})$.

Assume that $\underset{i\in\underline{k^{\prime\prime}-k^{\prime}}}{\bigcap}\text{{\bf null}}(L^{k^{\prime}+i-1})\neq\mathcal{R}$;
then there exists an $\boldsymbol{\eta}\notin\mathcal{R}$ such that
$L^{k^{\prime}+i-1}\boldsymbol{\eta}=\boldsymbol{0}$ for all $i\in\underline{k^{\prime\prime}-k^{\prime}}$,
which would imply,
\begin{align*}
\widetilde{L}_{[t_{k^{\prime}},t_{k^{\prime\prime}})}\boldsymbol{\eta} & =\left(\frac{1}{\triangle t_{k^{\prime}}}\intop_{t_{k^{\prime}}}^{t_{k^{\prime\prime}}}L(t)dt\right)\boldsymbol{\eta}\\
 & =\frac{1}{\triangle t_{k^{\prime}}}\sum_{i=1}^{k^{\prime\prime}-k^{\prime}}L^{k^{\prime}+i-1}(t_{k^{\prime}+i}-t_{k^{\prime}+i-1})\boldsymbol{\eta}\\
 & =\boldsymbol{0},
\end{align*}
contradicting the fact that $\text{{\bf null}}(\widetilde{L}_{[t_{k^{\prime}},t_{k^{\prime\prime}})})=\mathcal{R}$.
Therefore, $\underset{i\in\underline{k^{\prime\prime}-k^{\prime}}}{\bigcap}\text{{\bf null}}(L^{k^{\prime}+i-1})=\mathcal{R}$.

(Sufficiency) Assume that $\text{{\bf null}}(\widetilde{L}_{[t_{k^{\prime}},t_{k^{\prime\prime}})})\neq\mathcal{R}$;
then there exists $\boldsymbol{\eta}\notin\mathcal{R}$ such that
$\widetilde{L}_{[t_{k^{\prime}},t_{k^{\prime\prime}})}\boldsymbol{\eta}=\boldsymbol{0}$.
Hence, $\boldsymbol{\eta}^{\top}\widetilde{L}_{[t_{k^{\prime}},t_{k^{\prime\prime}})}\boldsymbol{\eta}=0$,
implying that,
\begin{align*}
 & \frac{1}{\triangle t_{k^{\prime}}}\boldsymbol{\eta}^{\top}\left(\intop_{t_{k^{\prime}}}^{t_{k^{\prime\prime}}}L(t)dt\right)\boldsymbol{\eta}\\
 & =\frac{1}{\triangle t_{k^{\prime}}}\sum_{i=1}^{k^{\prime\prime}-k^{\prime}}\boldsymbol{\eta}^{\top}L^{k^{\prime}+i-1}(t_{k^{\prime}+i}-t_{k^{\prime}+i-1})\boldsymbol{\eta}\\
 & =0.
\end{align*}
Due to the fact that $L^{k^{\prime}+i-1}$ is positive semi-definite
for all $i\in\underline{k^{\prime\prime}-k^{\prime}}$, $\boldsymbol{\eta}^{\top}L^{k^{\prime}+i-1}\boldsymbol{\eta}=0$,
which would imply that $L^{k^{\prime}+i-1}\boldsymbol{\eta}=\boldsymbol{0}$;
this on the other hand, contradicts the premise $\underset{i\in\underline{k^{\prime\prime}-k^{\prime}}}{\bigcap}\text{{\bf null}}(L^{k^{\prime}+i-1})=\mathcal{R}$.
Thus $\text{{\bf null}}(\widetilde{L}_{[t_{k^{\prime}},t_{k^{\prime\prime}})})=\mathcal{R}$.
\end{IEEEproof}
In order to link the state evolution of the multi-agent system \eqref{equ:matrix-consensus-overall}
and the null space of the integral of matrix-weighted time-varying
networks, we need to employ the state transition matrix. Denote $\varPhi(k^{\prime},k^{\prime\prime})=e^{-L^{k^{\prime\prime}-1}\triangle t_{k^{\prime\prime}-1}}\cdots e^{-L^{k^{\prime}}\triangle t_{k^{\prime}}}$.
Then $\boldsymbol{x}(t_{k^{\prime\prime}})=\varPhi(k^{\prime},k^{\prime\prime})\boldsymbol{x}(t_{k^{\prime}})$,
where $k^{\prime}<k^{\prime\prime}\in\mathbb{N}$. Note that the matrix-valued
Laplacian $L$ has at least $d$ zero eigenvalues. Let $\lambda_{1}\leq\lambda_{2}\leq\cdots\leq\lambda_{dn}$
be the eigenvalues of $L$. Then we have $0=\lambda_{1}=\cdots=\lambda_{d}\leq\lambda_{d+1}\le\cdots\leq\lambda_{dn}$.
Denote by $\beta_{1}\geq\beta_{2}\geq\cdots\geq\beta_{dn}$ as the
eigenvalues of $e^{-Lt}$; then $\beta_{i}(e^{-Lt})=e^{-\lambda_{i}(L)t}$,
i.e., $1=\beta_{1}=\cdots=\beta_{d}\geq\beta_{d+1}\geq\cdots\geq\beta_{dn}$.
In the meantime, the eigenvector corresponding to the eigenvalue
$\beta_{i}(e^{-Lt})$ is equal to that corresponding to $\lambda_{i}(L)$.
Consider the symmetric matrix $\varPhi(k^{\prime},k^{\prime\prime})^{\top}\varPhi(k^{\prime},k^{\prime\prime})$
which has at least $d$ eigenvalues at $1$. Let $\mu_{j}$ be the
eigenvalues of $\varPhi(k^{\prime},k^{\prime\prime})^{\top}\varPhi(k^{\prime},k^{\prime\prime})$,
where $j\in\underline{dn}$ such that $\mu_{1}=\cdots=\mu_{d}=1$
and $\mu_{d+1}\geq\mu_{d+2}\geq\cdots\geq\mu_{dn}$. The following
lemma provides the relationship between the null space of the matrix-valued
Laplacian of $\widetilde{\mathcal{G}}_{[t_{k^{\prime}},t_{k^{\prime\prime}})}$
and the eigenvalue $\mu_{d+1}$ of $\varPhi(k^{\prime},k^{\prime\prime})^{\top}\varPhi(k^{\prime},k^{\prime\prime})$. This relationship will prove useful in the proof of our main theorem.
\begin{lem}
Let $\mathcal{G}(t)$ be a matrix-weighted time-varying network satisfying
\textbf{Assumption 1. }Then $\text{{\bf null}}(\widetilde{L}_{[t_{k^{\prime}},t_{k^{\prime\prime}})})=\mathcal{R}$
if and only if 
\[
\mu_{d+1}(\varPhi(k^{\prime},k^{\prime\prime})^{\top}\varPhi(k^{\prime},k^{\prime\prime}))<1,
\]
where $k^{\prime}<k^{\prime\prime}\in\mathbb{N}$.
\end{lem}
\begin{IEEEproof}
(Sufficiency) Assume that $\text{{\bf null}}(\widetilde{L}_{[t_{k^{\prime}},t_{k^{\prime\prime}})})\neq\mathcal{R}$;
then according to Lemma \ref{thm:null space equality}, there exists
an $\boldsymbol{\eta}\notin\mathcal{R}$ such that $L^{k^{\prime}+i-1}\boldsymbol{\eta}=\boldsymbol{0}$
for all $i\in\underline{k^{\prime\prime}-k^{\prime}}$. Thus one can
obtain $e^{-L^{k^{\prime}+i-1}t}\boldsymbol{\eta}=\boldsymbol{\eta}$
for all $i\in\underline{k^{\prime\prime}-k^{\prime}}$ and $\varPhi(k^{\prime},k^{\prime\prime})\boldsymbol{\eta}=\boldsymbol{\eta}$.
According to the Lemma \ref{lem:Rayleigh Theorem}, one has
\begin{align*}
 & \mu_{d+1}(\varPhi(k^{\prime},k^{\prime\prime})^{\top}\varPhi(k^{\prime},k^{\prime\prime}))\\
\geq & \frac{\boldsymbol{\eta}^{\top}\varPhi(k^{\prime},k^{\prime\prime})^{\top}\varPhi(k^{\prime},k^{\prime\prime})\boldsymbol{\eta}}{\boldsymbol{\eta}^{\top}\boldsymbol{\eta}}\\
= & 1,
\end{align*}
contradicting,
\[
\mu_{d+1}(\varPhi(k^{\prime},k^{\prime\prime})^{\top}\varPhi(k^{\prime},k^{\prime\prime}))<1.
\]
Therefore $\text{{\bf null}}(\widetilde{L}_{[t_{k^{\prime}},t_{k^{\prime\prime}})})=\mathcal{R}$
holds.

(Necessity) Assume that $\mu_{d+1}(\varPhi(k^{\prime},k^{\prime\prime})^{\top}\varPhi(k^{\prime},k^{\prime\prime}))\geq1$.
Again, according to Lemma \ref{lem:Rayleigh Theorem}, there exists
a $\boldsymbol{\eta}\notin\mathcal{R}$ and $\boldsymbol{\eta}\neq\boldsymbol{0}$
such that
\begin{align*}
\mu_{d+1}(\varPhi(k^{\prime},k^{\prime\prime})^{\top}\varPhi(k^{\prime},k^{\prime\prime}))
& = \frac{\boldsymbol{\eta}^{\top}\varPhi(k^{\prime},k^{\prime\prime})^{\top}\varPhi(k^{\prime},k^{\prime\prime})\boldsymbol{\eta}}{\boldsymbol{\eta}^{\top}\boldsymbol{\eta}}\\
&\geq 1.
\end{align*}
Thus,
\[
\parallel\boldsymbol{\eta}\parallel\leq\parallel\varPhi(k^{\prime},k^{\prime\prime})\boldsymbol{\eta}\parallel.
\]

Let $\boldsymbol{\eta}_{k^{\prime}}=\boldsymbol{\eta}$ and $\boldsymbol{\eta}_{k^{\prime}+i}=e^{-L^{k^{\prime}+i-1}\triangle t_{k^{\prime}+i-1}}\boldsymbol{\eta}_{k^{\prime}+i-1}$
for $i\in\underline{k^{\prime\prime}-k^{\prime}}$. Due to the fact
$\lambda_{j}(e^{-L^{k^{\prime}+i-1}\triangle t_{k^{\prime}+i-1}})\leq1$
for $j\in\underline{dn}$ and $\boldsymbol{\eta}\notin\mathcal{R}$,
then
\[
\parallel e^{-L^{k^{\prime}+i-1}\triangle t_{k^{\prime}+i-1}}\boldsymbol{\eta}\parallel\leq\parallel\boldsymbol{\eta}\parallel,
\]
which implies that,
\begin{align*}
\parallel\boldsymbol{\eta}\parallel & \leq\parallel\varPhi(k^{\prime},k^{\prime\prime})\boldsymbol{\eta}\parallel\\
 & =\parallel\boldsymbol{\eta}_{k^{\prime\prime}}\parallel\leq\parallel\boldsymbol{\eta}_{k^{\prime\prime}-1}\parallel\leq\ldots\leq\parallel\boldsymbol{\eta}_{k^{\prime}}\parallel\\
 & =\parallel\boldsymbol{\eta}\parallel.
\end{align*}
Hence, $\parallel e^{-L^{k^{\prime}+i-1}\triangle t_{k^{\prime}+i-1}}\boldsymbol{\eta}_{k^{\prime}+i-1}\parallel=\parallel\boldsymbol{\eta}_{k^{\prime}+i-1}\parallel$
for $i\in\underline{k^{\prime\prime}-k^{\prime}}$. Then, one can
further derive $L^{k^{\prime}+i-1}\boldsymbol{\eta}_{k^{\prime}+i-1}=\boldsymbol{0}$;
thus $\boldsymbol{\eta}_{k^{\prime}+i-1}\in\text{{\bf ker}}(L^{k^{\prime}+i-1})$.
Note that since,
\begin{align*}
\parallel\boldsymbol{\eta}_{k^{\prime}+i}-\boldsymbol{\eta}_{k^{\prime}+i-1}\parallel & =\parallel e^{-L^{k^{\prime}+i-1}\triangle t_{k^{\prime}+i-1}}\boldsymbol{\eta}_{k^{\prime}+i-1}-\boldsymbol{\eta}_{k^{\prime}+i-1}\parallel\\
 & =\parallel\sum_{t=1}^{\infty}\frac{1}{t!}(-L^{k^{\prime}+i-1}\triangle t_{k^{\prime}+i-1})^{t}\boldsymbol{\eta}_{k^{\prime}+i-1}\parallel\\
 & =0,
\end{align*}
one can further obtain $\boldsymbol{\eta}_{k^{\prime}+i-1}=\boldsymbol{\eta}_{k^{\prime}+i}$
for $\forall i\in\underline{k^{\prime\prime}-k^{\prime}}$, which
implies that $\boldsymbol{\eta}\in\underset{i\in\underline{k^{\prime\prime}-k^{\prime}}}{\cap}\text{{\bf ker}}(L^{k^{\prime}+i-1})$
and $\text{{\bf null}}(\widetilde{L}_{[t_{k^{\prime}},t_{k^{\prime\prime}})})\neq\mathcal{R}$.
This is a contradiction however. As such $\mu_{d+1}(\varPhi(k^{\prime},k^{\prime\prime})^{\top}\varPhi(k^{\prime},k^{\prime\prime}))<1$.
\end{IEEEproof}
\begin{thm}
\label{thm:consensus theorem-necessary}Let $\mathcal{G}(t)$ be a
matrix-weighted time-varying network satisfying \textbf{Assumption
1.} If the multi-agent network \eqref{equ:matrix-consensus-overall}
admits an average consensus, then there exists a subsequence of $\{t_{k}|k\in\mathbb{N}\}$
denoted by $\{t_{k_{l}}|l\in\mathbb{N}\}$, such that the null space
of the matrix-valued Laplacian of $\widetilde{\mathcal{G}}_{[t_{k_{l}},t_{k_{l+1}})}(t)$
is $\mathcal{R}$, namely, $\text{{\bf null}}(\widetilde{L}_{[t_{k_{l}},t_{k_{l+1}})})=\mathcal{R}$
for all $l\in\mathbb{N}$, where $\triangle t_{k_{l}}=t_{k_{l+1}}-t_{k_{l}}<\infty$
and $t_{k_{0}}=t_{0}$.
\end{thm}
\begin{IEEEproof}
Assume that there does not exist a subsequence $\{t_{k_{l}}|l\in\mathbb{N}\}$
such that $\text{{\bf null}}(\widetilde{L}_{[t_{k_{l}},t_{k_{l+1}})})=\mathcal{R}$
for all $l\in\mathbb{N}$, which implies that there exists $k^{*}\in\mathbb{N}$
such that $\text{{\bf null}}(\widetilde{L}_{[t_{k^{*}},\infty)})\neq\mathcal{R}$.
Then $\underset{k\geq k^{*},k\in\mathbb{N}}{\bigcap}\text{{\bf null}}(L^{k})\neq\mathcal{R}$.
Denote $\boldsymbol{\eta}\notin\mathcal{R}$ and $\boldsymbol{\eta}\in\underset{k\geq k^{*},k\in\mathbb{N}}{\bigcap}\text{{\bf null}}(L^{k})$.
Then $L^{k}\boldsymbol{\eta}=\boldsymbol{0}$ for all $k\geq k^{*},k\in\mathbb{N}$.
One can choose a suitable $\boldsymbol{x}(0)$ such that $\boldsymbol{x}(t_{k^{*}})=\boldsymbol{\eta}$;
then $\underset{t\rightarrow\infty}{\text{{\bf lim}}}\boldsymbol{x}(t)=\boldsymbol{\eta}$,
establishing a contradiction to the fact that the multi-agent network
\eqref{equ:matrix-consensus-overall} admits an average consensus.
Thus, there exists a subsequence $\{t_{k_{l}}|l\in\mathbb{N}\}$
such that $\text{{\bf null}}(\widetilde{L}_{[t_{k_{l}},t_{k_{l+1}})})=\mathcal{R}$
for all $l\in\mathbb{N}$.
\end{IEEEproof}
\begin{rem}
Although the existence of a subsequence of $\{t_{k}|k\in\mathbb{N}\}$
denoted by $\{t_{k_{l}}|l\in\mathbb{N}\}$ such that $\text{{\bf null}}(\widetilde{L}_{[t_{k_{l}},t_{k_{l+1}})})=\mathcal{R}$,
for all $l\in\mathbb{N}$ is a necessary condition for an average
consensus, it is not sufficient. To see this fact, we choose, for
instance, the multi-agent system $\dot{\boldsymbol{x}}(t)=-\frac{1}{t^{2}}L\boldsymbol{x}(t)$,
where $L$ is the matrix-valued Laplacian of a time-invariant matrix-weighted
network for which $\text{{\bf null}}(L)=\mathcal{R}$. Now consider the
underlying matrix-weighted time-varying network corresponding to the
Laplacian matrix $\frac{1}{t^{2}}L$. Then for the arbitrary subsequence
$\{t_{k_{l}}|l\in\mathbb{N}\}$ of $\{t_{k}|k\in\mathbb{N}\}$, one
always has $\text{{\bf null}}(\widetilde{L}_{[t_{k_{l}},t_{k_{l+1}})})=\mathcal{R}$
for all $l\in\mathbb{N}$. However, the solution to the above system
is $\boldsymbol{x}(t)=e^{\frac{L}{t}}e^{-L}\boldsymbol{x}(0)$, and
$\text{{\bf lim}}_{t\rightarrow\infty}\boldsymbol{x}(t)=e^{-L}\boldsymbol{x}(0)$.
Therefore, an average consensus cannot be achieved in this example.
Thus, we need additional conditions in order to guarantee average consensus
for \eqref{equ:matrix-consensus-overall}. These observations motivate the following
result.
\end{rem}
\begin{thm}
\label{thm:consensus theorem}Let ${\normalcolor \mathcal{G}(t)}$
be a matrix-weighted time-varying network satisfying \textbf{Assumption
1}; furthermore, suppose there exists a subsequence of $\{t_{k}|k\in\mathbb{N}\}$, denoted
by $\{t_{k_{l}}|l\in\mathbb{N}\}$, such that $\text{{\bf null}}(\widetilde{L}_{[t_{k_{l}},t_{k_{l+1}})})=\mathcal{R}$
for all $l\in\mathbb{N}$, where $\triangle t_{k_{l}}=t_{k_{l+1}}-t_{k_{l}}<\infty$
and $t_{k_{0}}=t_{0}$. If there exists a scalar $0<q<1$
such that $\mu_{d+1}(\varPhi(t_{k_{l}},t_{k_{l+1}})^{\top}\varPhi(t_{k_{l}},t_{k_{l+1}}))\leq q$
for all $l\in\mathbb{N}$, then the multi-agent network \eqref{equ:matrix-consensus-overall}
admits an average consensus.
\end{thm}
\begin{IEEEproof}
Let $\boldsymbol{\omega}(t)=\boldsymbol{x}(t)-\boldsymbol{x}_{f}$.
Then $\dot{\boldsymbol{\omega}}(t)=-L(t)\boldsymbol{\omega}(t)$.
Choose $\boldsymbol{\omega}(0)\notin\mathcal{R}$ and observe that,
\begin{align*}
 & \mu_{d+1}(\varPhi(t_{k_{0}},t_{k_{1}})^{\top}\varPhi(t_{k_{0}},t_{k_{1}}))\\
\geq & \frac{\boldsymbol{\omega}(0)^{\top}(\varPhi(t_{k_{0}},t_{k_{1}})^{\top}\varPhi(t_{k_{0}},t_{k_{1}}))\boldsymbol{\omega}(0)}{\boldsymbol{\omega}(0)^{\top}\boldsymbol{\omega}(0)}\\
= & \frac{\boldsymbol{\omega}(t_{k_{1}})^{\top}\boldsymbol{\omega}(t_{k_{1}})}{\boldsymbol{\omega}(0)^{\top}\boldsymbol{\omega}(0)},
\end{align*}
implying that, 
\[
\parallel\boldsymbol{\omega}(t_{k_{1}})\parallel\leq\mu_{d+1}(\varPhi(t_{k_{0}},t_{k_{1}})^{\top}\varPhi(t_{k_{0}},t_{k_{1}}))^{\frac{1}{2}}\parallel\boldsymbol{\omega}(0)\parallel.
\]
Therefore, 
\begin{align*}
\parallel\boldsymbol{\omega}(t_{k_{l+1}})\parallel & \leq\mu_{d+1}(\varPhi(t_{k_{l}},t_{k_{l+1}})^{\top}\varPhi(t_{k_{l}},t_{k_{l+1}}))^{\frac{1}{2}}\parallel\boldsymbol{\omega}(t_{k_{l}})\parallel\\
 & \leq\mu_{d+1}(\varPhi(t_{k_{l}},t_{k_{l+1}})^{\top}\varPhi(t_{k_{l}},t_{k_{l+1}}))^{\frac{1}{2}}\\
{\color{red}} & \thinspace\thinspace\thinspace\thinspace\thinspace\thinspace\thinspace\thinspace\thinspace\thinspace\thinspace\thinspace\thinspace\thinspace\thinspace\thinspace\thinspace\thinspace\thinspace\thinspace\thinspace\thinspace\thinspace\thinspace\thinspace\thinspace\thinspace\thinspace\thinspace\thinspace\thinspace\thinspace\thinspace\thinspace\thinspace\thinspace\thinspace\thinspace\thinspace\thinspace\thinspace\thinspace\thinspace\thinspace\thinspace\thinspace\thinspace\thinspace\thinspace\thinspace\thinspace\vdots\\
{\color{red}} & \mu_{d+1}(\varPhi(t_{k_{0}},t_{k_{1}})^{\top}\varPhi(t_{k_{0}},t_{k_{1}}))^{\frac{1}{2}}\parallel\boldsymbol{\omega}(0)\parallel\\
 & \leq q^{\frac{1}{2}(l+1)}\parallel\boldsymbol{\omega}(0)\parallel.
\end{align*}
Let 
\[
V(t)=\boldsymbol{\omega}(t)^{\top}\boldsymbol{\omega}(t)=\parallel\boldsymbol{\omega}(t)\parallel^{2};
\]
then 
\[
\dot{V}(t)=2\boldsymbol{\omega}(t)^{\top}(-L(t))\boldsymbol{\omega}(t)\leq0.
\]
 Thus
\[
\parallel\boldsymbol{\omega}(t)\parallel\leq\parallel\boldsymbol{\omega}(t_{k_{l+1}})\parallel\leq q^{\frac{1}{2}(l+1)}\parallel\boldsymbol{\omega}(0)\parallel,
\]
for $\forall t\in[t_{k_{l+1}},\infty)$. Note that $0<q<1$, and hence,
\[
{\displaystyle \lim_{t\rightarrow\infty}}\parallel\boldsymbol{\omega}(t)\parallel=0.
\]
As such, the multi-agent network \eqref{equ:matrix-consensus-overall}
achieves average consensus.
\end{IEEEproof}

\section{Consensus on Periodic Matrix-weighted Time-varying Networks \label{sec:Consensus-on-Periodic}}

In the subsequent discussion, we consider a special class of time-varying
networks, where $\mathcal{G}(t)$ is periodic. The periodic network
$\mathcal{G}(t)$ is formally characterized by the following assumption.

\textbf{Assumption 2. } There exists a $T>0$ such that $\mathcal{G}(t+T)=\mathcal{G}(t)$
for any $t\geq0$. Moreover, there exists a time sequence $\left\{ t_{k}|k\in\mathbb{N}\right\}$
satisfying $\triangle t_{k}=t_{k+1}-t_{k}>\alpha$ for all $k\in\mathbb{N}$,
where $\alpha>0$, and there exists $m>2$ ($m\in\mathbb{N}$) partitions
for each time span $[lT,(l+1)T)$ for which,
\[
lT=t_{lm}<t_{lm+1}<\cdots<t_{(l+1)m}=(l+1)T,\thinspace l\in\mathbb{N},
\]
and $\mathcal{G}(t)$ is time-invariant for $t\in[t_{k},t_{k+1})$,
where $k\in\mathbb{N}$.

Under Assumption 2, we now proceed to provide the algebraic and graph-theoretic
conditions under which the multi-agent system \eqref{equ:matrix-consensus-overall}
admits average consensus.
\begin{thm}
\label{thm:consensus theorem-periodic}Let $\mathcal{G}(t)$ be a
periodic matrix-weighted time-varying network satisfying \textbf{Assumption
2}. Then the multi-agent network \eqref{equ:matrix-consensus-overall}
admits average consensus if and only if,
\[
\text{{\bf null}}(\widetilde{L}_{[0,T)})=\mathcal{R}.
\]
\end{thm}
\begin{IEEEproof}
(Necessity) Assume that $\text{{\bf null}}(\widetilde{L}_{[0,T)})\neq\mathcal{R}$;
then there exists a $\boldsymbol{\eta}\notin\mathcal{R}$ such that
$L^{i-1}\boldsymbol{\eta}=\boldsymbol{0}$ for all $i\in\underline{m}$.
Let $\boldsymbol{x}(0)=\boldsymbol{\eta}$. Thereby, we can obtain $\boldsymbol{x}(t)=\boldsymbol{\eta}$
for all $t>0$, contradicting the fact that the multi-agent
network \eqref{equ:matrix-consensus-overall} admits average consensus.

(Sufficiency) Let $\boldsymbol{\omega}(t)=\boldsymbol{x}(t)-\boldsymbol{x}_{f}$;
then we have $\dot{\boldsymbol{\omega}}(t)=-L(t)\boldsymbol{\omega}(t)$.
Denote 
\[
\varPhi(0,T)=e^{-L^{m-1}\triangle t_{m-1}}\cdots e^{-L^{0}\triangle t_{0}},
\]
and choose $\boldsymbol{\omega}(0)\notin\mathcal{R}$. Then,
\begin{align*}
\mu_{d+1}(\varPhi(0,T)^{\top}\varPhi(0,T)) & \geq\frac{\boldsymbol{\omega}(0)^{\top}(\varPhi(0,T)^{\top}\varPhi(0,T))\boldsymbol{\omega}(0)}{\boldsymbol{\omega}(0)^{\top}\boldsymbol{\omega}(0)}\\
 & =\frac{\boldsymbol{\omega}(T)^{\top}\boldsymbol{\omega}(T)}{\boldsymbol{\omega}(0)^{\top}\boldsymbol{\omega}(0)},
\end{align*}
implying that,
\[
\boldsymbol{\omega}(T)^{\top}\boldsymbol{\omega}(T)\leq\mu_{d+1}(\varPhi(0,T)^{\top}\varPhi(0,T))\boldsymbol{\omega}(0)^{\top}\boldsymbol{\omega}(0).
\]
Therefore
\[
\parallel\boldsymbol{\omega}(T)\parallel\leq\mu_{d+1}(\varPhi(0,T)^{\top}\varPhi(0,T))^{\frac{1}{2}}\parallel\boldsymbol{\omega}(0)\parallel,
\]
implying that,
\[
\parallel\boldsymbol{\omega}(kT)\parallel\leq\mu_{d+1}(\varPhi(0,T)^{\top}\varPhi(0,T))^{\frac{1}{2}k}\parallel\boldsymbol{\omega}(0)\parallel.
\]
Hence, one has
\[
\parallel\boldsymbol{\omega}(t)\parallel\leq\parallel\boldsymbol{\omega}(kT)\parallel\leq\mu_{d+1}(\varPhi(0,T)^{\top}\varPhi(0,T))^{\frac{1}{2}k}\parallel\boldsymbol{\omega}(0)\parallel,
\]
for $t\in[kT,(k+1)T)$; then $\text{{\bf lim}}_{t\rightarrow\infty}\parallel\boldsymbol{\omega}(t)\parallel=0$.
Therefore, the multi-agent network \eqref{equ:matrix-consensus-overall}
admits average consensus.
\end{IEEEproof}
Theorem \ref{thm:consensus theorem-periodic} provides
an algebraic condition for reaching consensus for periodic matrix-weighted
time-varying networks using the structure of the null space
of the matrix-valued Laplacian matrix of the corresponding integral
network. An analogous graph theoretic condition is as follows.
\begin{thm}
\label{thm:graph condition-integral graph}Let $\mathcal{G}(t)$ be
a periodic matrix-weighted time-varying network satisfying \textbf{Assumption
2}. If the integral graph of $\mathcal{G}(t)$ over time span $[0,T)$
has a positive spanning tree, then the multi-agent network \eqref{equ:matrix-consensus-overall}
admits average consensus.
\end{thm}
\begin{IEEEproof}
Let $\widetilde{\mathcal{G}}_{[0,T)}$ be the integral network of
$\mathcal{G}(t)$ over time span $[0,T)$. If $\widetilde{\mathcal{G}}_{[0,T)}$
has a positive spanning tree, from Lemma \ref{lem:2} and Lemma \ref{lem:3},
one has $\text{{\bf null}}(\widetilde{L}_{[0,T)})=\mathcal{R}$, where
$\widetilde{L}_{[0,T)}$ is the matrix-valued Laplacian matrix of $\widetilde{\mathcal{G}}_{[0,T)}$.
Theorem \ref{thm:consensus theorem-periodic}, now implies that
the multi-agent network \eqref{equ:matrix-consensus-overall}
admits average consensus.
\end{IEEEproof}

\section{Simulation Results \label{sec:Simulation-Results}}

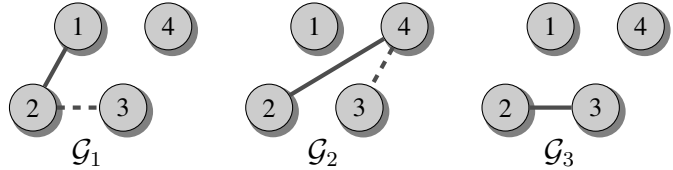
\begin{figure}[H]
\begin{centering}
\begin{tikzpicture}[scale=0.6]
	
    \node (n1) at (1 ,0.3) [circle,circular drop shadow,fill=black!20,draw] {1};
	\node (n2) at (0,-1.5) [circle,circular drop shadow,fill=black!20,draw] {2};
    \node (n3) at (2,-1.5) [circle,circular drop shadow,fill=black!20,draw] {3};
    \node (n4) at (3 ,0.3) [circle,circular drop shadow,fill=black!20,draw] {4};
   
    \node (G2) at (1.2,-2.5) {\large{$\mathcal{G}_1$}};



	\draw[-, ultra thick, color=black!70] [-] (n1) -- (n2); 
	\draw [-, ultra thick, dashed, color=black!70] (n2) -- (n3); 

\end{tikzpicture}\,\,\,\,\,\,\,\,\,\,\,\,\begin{tikzpicture}[scale=0.6]

    \node (n1) at (1 ,0.3) [circle,circular drop shadow,fill=black!20,draw] {1};
	\node (n2) at (0,-1.5) [circle,circular drop shadow,fill=black!20,draw] {2};
    \node (n3) at (2,-1.5) [circle,circular drop shadow,fill=black!20,draw] {3};
    \node (n4) at (3 ,0.3) [circle,circular drop shadow,fill=black!20,draw] {4};
    \node (G2) at (1.2,-2.5) {\large{$\mathcal{G}_2$}};

	\draw[-, ultra thick, color=black!70] (n2) -- (n4); 
	\draw[-, ultra thick, dashed, color=black!70] (n4) -- (n3); 
	
\end{tikzpicture}\,\,\,\,\,\,\,\,\,\,\,\,\begin{tikzpicture}[scale=0.6]

    \node (n1) at (1 ,0.3) [circle,circular drop shadow,fill=black!20,draw] {1};
	\node (n2) at (0,-1.5) [circle,circular drop shadow,fill=black!20,draw] {2};
    \node (n3) at (2,-1.5) [circle,circular drop shadow,fill=black!20,draw] {3};
    \node (n4) at (3 ,0.3) [circle,circular drop shadow,fill=black!20,draw] {4};
    
    \node (G2) at (1.2,-2.5) {\large{$\mathcal{G}_3$}};

	\draw[-, ultra thick, color=black!70] (n2) -- (n3); 
	
\end{tikzpicture}
\par\end{centering}
\caption{Three matrix-weighted networks $\mathcal{G}_{1}$, $\mathcal{G}_{2}$,
and $\mathcal{G}_{3}$. \textcolor{black}{Those edges weighted by
positive definite matrices are illustrated by solid lines and edges
weighted by positive semi-definite matrices are illustrated by dotted
lines.}}
\label{fig:example-network}
\end{figure}

Consider a sequence of matrix-weighted networks, consisting 
of (the same) four agents, and the topologies of the networks are as $\mathcal{G}_{1},\mathcal{G}_{2}$
and $\mathcal{G}_{3}$, as shown in Figure \ref{fig:example-network}.
Note that $n=4$ and $d=2$ in this example.

The matrix-valued edge weights for each network are,
\[
A_{12}(\mathcal{G}_{1})=\left[\begin{array}{cc}
1 & 1\\
1 & 2
\end{array}\right],\thinspace\thinspace A_{23}(\mathcal{G}_{1})=\left[\begin{array}{cc}
1 & 1\\
1 & 1
\end{array}\right],
\]
\[
A_{24}(\mathcal{G}_{2})=\left[\begin{array}{cc}
1 & 0\\
0 & 2
\end{array}\right],\thinspace\thinspace A_{34}(\mathcal{G}_{2})=\left[\begin{array}{cc}
1 & 0\\
0 & 0
\end{array}\right],
\]
and
\[
A_{23}(\mathcal{G}_{3})=\left[\begin{array}{cc}
1 & -1\\
-1 & 2
\end{array}\right],
\]
respectively. The matrix-valued Laplacian matrices corresponding to
above three networks are, 

\[
L(\mathcal{G}_{1})=\left[\begin{array}{cccccccc}
1 & 1 & -1 & -1 & 0 & 0 & 0 & 0\\
1 & 2 & -1 & -2 & 0 & 0 & 0 & 0\\
-1 & -1 & 2 & 2 & -1 & -1 & 0 & 0\\
-1 & -2 & 2 & 3 & -1 & -1 & 0 & 0\\
0 & 0 & -1 & -1 & 1 & 1 & 0 & 0\\
0 & 0 & -1 & -1 & 1 & 1 & 0 & 0\\
0 & 0 & 0 & 0 & 0 & 0 & 0 & 0\\
0 & 0 & 0 & 0 & 0 & 0 & 0 & 0
\end{array}\right],
\]

\[
L(\mathcal{G}_{2})=\left[\begin{array}{cccccccc}
0 & 0 & 0 & 0 & 0 & 0 & 0 & 0\\
0 & 0 & 0 & 0 & 0 & 0 & 0 & 0\\
0 & 0 & 1 & 0 & 0 & 0 & -1 & 0\\
0 & 0 & 0 & 2 & 0 & 0 & 0 & -2\\
0 & 0 & 0 & 0 & 1 & 0 & -1 & 0\\
0 & 0 & 0 & 0 & 0 & 0 & 0 & 0\\
0 & 0 & -1 & 0 & -1 & 0 & 2 & 0\\
0 & 0 & 0 & -2 & 0 & 0 & 0 & 2
\end{array}\right],
\]
and

\[
L(\mathcal{G}_{3})=\left[\begin{array}{cccccccc}
0 & 0 & 0 & 0 & 0 & 0 & 0 & 0\\
0 & 0 & 0 & 0 & 0 & 0 & 0 & 0\\
0 & 0 & 1 & -1 & -1 & 1 & 0 & 0\\
0 & 0 & -1 & 2 & 1 & -2 & 0 & 0\\
0 & 0 & -1 & 1 & 1 & -1 & 0 & 0\\
0 & 0 & 1 & -2 & -1 & 2 & 0 & 0\\
0 & 0 & 0 & 0 & 0 & 0 & 0 & 0\\
0 & 0 & 0 & 0 & 0 & 0 & 0 & 0
\end{array}\right],
\]
respectively.

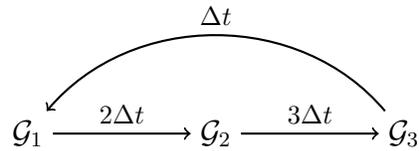
\begin{figure}[tbh]
\begin{centering}
\begin{tikzpicture}[scale=1]

    \node (G1) at (0,0) [] {\large{$\mathcal{G}_1$}};
    \node (G2) at (2.5,0) [] {\large{$\mathcal{G}_2$}};
    \node (G3) at (5,0) [] {\large{$\mathcal{G}_3$}};





\path[]
	(G1) [->,  thick] edge node[above] {$ 2\Delta t$} (G2);

\path[]
	(G2) [->,  thick]  edge node[above] {$3\Delta t$} (G3);

\path[]
	(G1) [<-,  thick, bend left=50] edge node[above] {$\Delta t$} (G3);

\end{tikzpicture}
\par\end{centering}
\caption{Switching sequence amongst networks $\mathcal{G}_{1}$, $\mathcal{G}_{2}$,
and $\mathcal{G}_{3}$.}
\label{fig:switching-network}
\end{figure}

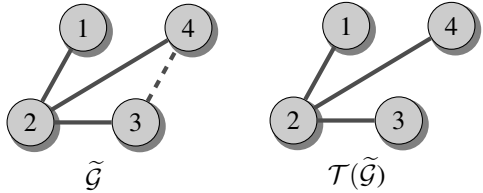
\begin{figure}[tbh]
\begin{centering}
\begin{tikzpicture}[scale=0.7]

    \node (n1) at (1 ,0.3) [circle,circular drop shadow,fill=black!20,draw] {1};
	\node (n2) at (0,-1.5) [circle,circular drop shadow,fill=black!20,draw] {2};
    \node (n3) at (2,-1.5) [circle,circular drop shadow,fill=black!20,draw] {3};
    \node (n4) at (3 ,0.3) [circle,circular drop shadow,fill=black!20,draw] {4};
   
    \node (G2) at (1.2,-2.5) {{$\widetilde{\mathcal{G}}$}};


	\draw[-, ultra thick, color=black!70] (n1) -- (n2); 
	\draw[-, ultra thick, color=black!70] (n2) -- (n3); 
    \draw[-, ultra thick, color=black!70] (n2) -- (n4); 
    \draw[-, ultra thick, dashed, color=black!70] (n3) -- (n4); 

\end{tikzpicture}\,\,\,\,\,\,\,\,\,\,\,\,\,\begin{tikzpicture}[scale=0.7]

    \node (n1) at (1 ,0.3) [circle,circular drop shadow,fill=black!20,draw] {1};
	\node (n2) at (0,-1.5) [circle,circular drop shadow,fill=black!20,draw] {2};
    \node (n3) at (2,-1.5) [circle,circular drop shadow,fill=black!20,draw] {3};
    \node (n4) at (3 ,0.3) [circle,circular drop shadow,fill=black!20,draw] {4};
   
    \node (G2) at (1.2,-2.5) {{$\mathcal{T}(\widetilde{\mathcal{G}}$})};



	\draw[-, ultra thick, color=black!70] (n1) -- (n2); 
	\draw[-, ultra thick, color=black!70] (n2) -- (n3); 
    \draw[-, ultra thick, color=black!70] (n2) -- (n4);

\end{tikzpicture}
\par\end{centering}
\caption{The integral graph $\mathcal{G}(t)$ over time span $[t_{6l},t_{6(l+1)})$
where $l\in\mathbb{N}$ (left) and the associated positive spanning
tree $\mathcal{T}(\widetilde{\mathcal{G}})$ (right).}
\label{fig:integral-network}
\end{figure}

\begin{figure}[tbh]
\begin{centering}
\includegraphics[width=9cm]{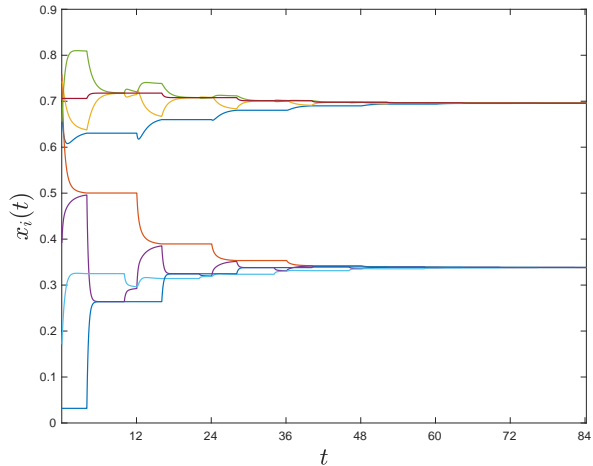}
\par\end{centering}
\caption{State evolution in the multi-agent system \eqref{equ:matrix-consensus-overall}.}
\label{fig:trajectory}
\end{figure}

Consider a time sequence $\{t_{k}\thinspace|\thinspace k\in\mathbb{N}\}$
such that $t_{k}=k\Delta t$ where $\Delta t>0$. The evolution is
initiated from network $\mathcal{G}_{1}$ (i.e., ${\normalcolor \mathcal{G}(0)=\mathcal{G}_{1}}$)
with $\boldsymbol{x}_{1}(0)=[0.6787,\thinspace0.7577]^{\top}$, $\boldsymbol{x}_{2}(0)=[0.7431,\thinspace0.3922]^{\top}$,
$\boldsymbol{x}_{3}(0)=[0.6555,\thinspace0.1712]^{\top}$ and $\boldsymbol{x}_{4}(0)=[0.7060,\thinspace0.0318]^{\top}$.
The switching among networks $\mathcal{G}_{1},\mathcal{G}_{2}$ and
$\mathcal{G}_{3}$ satisfies $\{t_{k}\thinspace|\thinspace k\in\mathbb{N}\}$,
\[
{\normalcolor \mathcal{G}(t)}=\begin{cases}
\begin{array}{c}
\mathcal{G}_{1},\\
\mathcal{G}_{2},\\
\mathcal{G}_{3},
\end{array} & \begin{array}{c}
t\in[t_{6l},t_{6l+2}),\\
t\in[t_{6l+2},t_{6l+5}),\\
t\in[t_{6l+5},t_{6(l+1)}),
\end{array}\end{cases}
\]
where $l\in\mathbb{N}$. The network switching process is demonstrated
in Figure \ref{fig:switching-network}. Examine the dimension of the
null space of $L(\mathcal{G}_{1})$, $L(\mathcal{G}_{2})$ and $L(\mathcal{G}_{3})$,
respectively. We have $\text{{\bf null}}(L(\mathcal{G}_{1}))\not=\mathcal{R}$,
$\text{{\bf null}}(L(\mathcal{G}_{2}))\not=\mathcal{R}$ and $\text{{\bf null}}(L(\mathcal{G}_{3}))\not=\mathcal{R}$.
However, note that from Figure \ref{fig:integral-network}, 
the integral graph of $L(\mathcal{G}_{1})$, $L(\mathcal{G}_{2})$
and $L(\mathcal{G}_{3})$ over time span $[t_{6l},t_{6(l+1)})$, where
$l\in\mathbb{N}$, denoted by $\widetilde{\mathcal{G}}$, has a positive
spanning tree $\mathcal{T}(\widetilde{\mathcal{G}})$. Therefore,
according to Theorem \ref{thm:graph condition-integral graph}, the
multi-agent system \eqref{equ:matrix-consensus-overall} admits an
average consensus solution at $[0.6958,\thinspace0.3382]^{\top}$;
see Figure \ref{fig:trajectory}.

\section{Conclusion \label{sec:Conclusion}}

This paper examines consensus problems on matrix-weighted time-varying
networks. For such networks, necessary
and/or sufficient conditions for reaching average consensus are provided. 
Furthermore, for matrix-weighted
periodic time-varying networks, necessary and sufficient algebraic 
and graph theoretic conditions are obtained for reaching consensus.

\bibliographystyle{IEEEtran}
\phantomsection\addcontentsline{toc}{section}{\refname}\bibliography{mybib}

\end{document}